\documentclass[sigconf]{acmart}
\acmJournal{TOSEM}
\usepackage{tikz}
\usepackage{amsmath}
\usepackage{enumitem}
\usepackage{algorithm}
\usepackage{algpseudocode}
\usepackage{graphicx}
\usepackage{caption}
\usepackage{subcaption}
\usepackage{booktabs}
\usepackage{xcolor}
\usepackage{multirow}
\usepackage{makecell}
\usepackage{tcolorbox}
\usepackage{array}

\newcommand{\tool}{\textsc{VLTest}}

\newcommand{\cf}{\mathrm{CF}}
\newcommand{\vjf}{\mathrm{AJF}}
\newcommand{\cfk}{\mathrm{CF@1k}}
\newcommand{\vjfk}{\mathrm{AJF@1k}}
\newcommand{\norm}{\mathrm{norm}}
\newcommand{\call}{\textsc{Call}}
\newcolumntype{L}[1]{>{\raggedright\arraybackslash}p{#1}}

\begin{document}

\date{}

\title{\tool: Finding Joint Image--Text Failures in Vision--Language Models}

\author{Bin Duan}
\affiliation{%
  \institution{The University of Queensland}
  \city{Brisbane}
  \country{Australia}}
\email{b.duan@uq.edu.au}

\author{Jintao Lin}
\affiliation{%
  \institution{The University of Queensland}
  \city{Brisbane}
  \country{Australia}}
\email{jintao.lin@student.uq.edu.au}

\author{Guowei Yang}
\authornote{Corresponding author.}
\affiliation{%
  \institution{The University of Queensland}
  \city{Brisbane}
  \country{Australia}}
\email{guowei.yang@uq.edu.au}

\renewcommand{\shortauthors}{Duan et al.}

\begin{abstract}
Vision--language models (VLMs) are increasingly embedded in software systems. Testing them raises two problems. Mutating either input can change the expected answer, so an output change is not by itself a failure. Some failures appear only when both inputs are mutated; attributing them to the combination requires testing each mutation alone. We present \tool, a black-box VLM testing approach. It generates image variants by editing VQGAN latent codes and text variants with a rewrite--expand--prune pipeline, filters variants likely to change the expected answer, and tests each variant alone before testing combinations, so every joint failure has four matched cells. A VLM judge confirms a failure only after checking that the input keeps the expected answer. On three tasks with eight VLMs, including two commercial API models, the oracle confirms 81.7\% of the candidates, with 88.8\% to 94.6\% precision against human labels, and four-cell confirmation rejects 22.5\% of joint candidates. Under the same budget, \tool's generators find 1.8 to 2.5 times as many confirmed failures as SINVAD and TextFooler, and its joint tests find 91 to 197 attributed joint failures per 1,000 joint calls, 1.3 times as many as the same pairing applied to the baselines' variants.
\end{abstract}

\maketitle

\section{Introduction}
\label{sec:intro}

Vision--language models (VLMs)~\cite{zhang2024vision,radford2021learning,li2023blip} take an image and a text as input and produce a decision or a natural-language answer. They power image--text matching~\cite{fu2023learning}, visual reasoning~\cite{gupta2023visual}, visual question answering~\cite{lu2023multi}, cross-modal retrieval~\cite{wang2025cross}, and multimodal search~\cite{barbany2024leveraging}, and they are increasingly integrated into software systems~\cite{aflalo2022vl}. Their predictions can change under small variations of either input~\cite{sim2025can,chang2024language,chen2023benchmarking}, which makes their behavioral reliability a software quality concern.

Software engineering research has developed effective techniques for testing deep neural networks (DNNs). Seed-centric generators explore the neighborhood of an input under a budget~\cite{pei2017deepxplore,guo2018dlfuzz,lee2020effective,wang2022bet,duan2025generating,duan2026latent,duan2026testing}, latent-space generators keep the generated inputs realistic~\cite{kang2020sinvad,dola2024cit4dnn,weissl2025mimicry}, and metamorphic relations provide test oracles when the expected output of a generated input is unknown~\cite{chen2018metamorphic,segura2016survey}. Applying these techniques to VLMs, however, raises two challenges.

\textbf{Challenge 1: output changes are not failures.} Seed-centric testing usually relies on an invariance relation: if a mutation preserves the content that determines the answer, the answer should not change. In a VLM, the expected answer depends on how the image and the text relate. A paraphrase can change what a question asks, a visual edit can change whether a statement is true, and two edits that are harmless on their own can change the answer together. An image--caption pair can remain well aligned while its correct match label changes. Counting output changes, as robustness evaluations of VLMs commonly do~\cite{chen2023benchmarking}, therefore mixes real failures with correct reactions to changed inputs. Studies of label preservation for generated test inputs focus on unimodal classifiers~\cite{maryam2025benchmarking,maryam2026deepnaqqal}; for VLMs, the expected answer also depends on the relation between the two inputs.

\textbf{Challenge 2: joint failures need attribution.} Some failures occur only when both inputs are mutated. To show that such a failure is caused by the combination, a tester must also run the image-only and the text-only variants, check that the model handles both correctly, and check that all inputs share the same expected answer. We call these four inputs the \emph{four cells} of a joint test: the seed, the image-only input, the text-only input, and the joint input. The cells should differ only by the mutations they contain; an image mutation made in the latent space of a generative model also carries reconstruction artifacts, which the text-only input must then share. Even with matched controls, the effects of two mutations can add up to a failure without any interaction between the modalities~\cite{hessel2020does}. Existing approaches that mutate both modalities, such as metamorphic testing of VQA systems~\cite{xu2026metara} and multimodal adversarial attacks~\cite{zhang2022towards,lu2023set,yin2023vlattack}, do not pair the final joint input with matched single-modality controls whose expected answers are confirmed by an oracle.
Without such controls, a higher failure count under joint mutation may simply reflect that two sources of perturbation were used instead of one.

To address these challenges, we propose \tool, a black-box testing approach for VLMs. The key idea of \tool\ is to treat a joint input and its two single-modality counterparts as one test, and to report a failure only after its expected answer has been confirmed by an oracle. \tool\ has three components. First, two controlled generators produce visual variants, by replacing dominant codebook entries in the discrete latent space of a pretrained VQGAN~\cite{esser2021taming}, and textual variants, through a rewrite--expand--prune pipeline. Both emit deduplicated variants in a fixed order. Second, relation filters remove variants that are likely to change the expected answer, using bidirectional entailment for text, an image similarity floor, and task-specific image--text alignment rules. Third, a two-phase testing procedure admits only seeds that the model answers correctly, first tests each variant on its own, and then tests combinations of these variants. All four cells are built on the reconstructed seed image, so that each cell differs from the seed only by the mutations it contains. A feedback-directed pairing strategy uses the single-modality results to test first the combinations that single-modality testing has not already broken.

\tool\ confirms failures with an automated oracle. A VLM judge receives the seed and its answer as a reference and first decides, without seeing the model output, whether a generated input keeps that answer; only then does it decide whether the output is wrong. A failure is reported only if both checks succeed, and a joint failure only if they succeed for all four cells. Human labels on a stratified sample of the reported failures measure the precision of the oracle.

We evaluate \tool\ on image--text matching (MSCOCO), visual reasoning (NLVR2), and visual question answering (VQA~v2) with eight VLMs: four task-specific models with six fine-tuned checkpoints, Qwen3-VL at two scales, and two commercial API models. We compare the generators of \tool\ with SINVAD~\cite{kang2020sinvad}, adapted to the VQGAN latent, and TextFooler~\cite{jin2020bert}, and its joint testing with five baselines: random pairing with and without filtering, random pairing that first uses the combinations whose single-modality inputs are answered correctly, stable-first pairing of the variants found by SINVAD and TextFooler, and the applicable relations of MetaRA~\cite{xu2026metara}. The oracle confirms 81.7\% of the candidates of \tool\ (SF) and reaches a precision of 80.0\% to 94.6\% against human labels across the compared methods in the primary experiment, and of 88.8\% to 94.6\% on the failures that \tool\ (SF) reports; four-cell confirmation rejects 22.5\% of the joint candidates, 19.7\% through the joint input alone and a further 2.8\% through the single-modality controls. Under the same budget, the generators of \tool\ find 1.8 to 2.5 times as many confirmed failures as SINVAD and TextFooler, and \tool\ finds 91 to 197 attributed joint failures per 1,000 joint calls, 1.3 times as many as the stable-first pairing of their variants and 2.8 times as many as the applicable relations of MetaRA, which exhaust their combinations after about half of the joint budget. Under an additive margin model, 16.5\% of the attributed joint failures with output scores are interactions between the modalities that the sum of the two single-modality effects does not predict. On Qwen3-VL, GPT-4o mini, and Gemini 2.5 Flash, \tool\ finds 112 to 250 attributed joint failures per 1,000 joint calls.

The main contributions of this paper are as follows.
\begin{itemize}[leftmargin=*]
\item We formulate VLM testing with oracle-confirmed expected answers and introduce four-cell evidence, built on a common reconstructed reference, for attributing a failure to the combination of an image mutation and a text mutation.
\item We propose \tool, which combines controlled visual and textual generators, relation filters, and two-phase testing that reuses the single-modality tests as the matched controls of every joint test and as feedback for pairing.
\item We design an automated test oracle that, for every cell, checks the expected answer against the seed with a VLM judge before it assesses the model output, and we validate the oracle against human labels.
\item We provide a replication artifact~\cite{Our} for \tool.
\end{itemize}

\section{Background and Related Work}
\label{sec:related}

\subsection{Background}
VLMs encode an image and a text into a shared or fused representation and decode a task output~\cite{zhang2024vision,manzoor2023multimodality}. Architectures range from single-stream fusion encoders such as ViLT~\cite{kim2021vilt}, to encoders pretrained at scale such as BLIP~\cite{li2022blip} and BEiT-3~\cite{wang2023image}, to generative models that couple a vision encoder with a language model, such as PaliGemma~\cite{beyer2024paligemma} and recent open-weight families~\cite{bai2025qwen25vl,zhu2025internvl3,li2024llavaonevision,qwen2025qwen3vl}. We consider three tasks with different output types: image--text matching (ITM, binary), visual reasoning over a statement and an image pair (VR, binary), and visual question answering (VQA, short free-form answers with several acceptable answers).

\subsection{Related Work}

Table~\ref{tab:paradigms} compares \tool\ with representative approaches.

\noindent\textbf{Test Input Generation for DNNs.}
Seed-centric test generation explores the neighborhood of an input to expose behaviors that fixed datasets miss~\cite{sun2018testing}. Coverage-guided methods perturb inputs in pixel space~\cite{pei2017deepxplore,guo2018dlfuzz,lee2020effective,wang2022bet}. Generative methods search a learned latent space so that the generated inputs remain realistic: SINVAD~\cite{kang2020sinvad} searches the latent space of a VAE, CIT4DNN~\cite{dola2024cit4dnn} applies combinatorial coverage to a discretized latent space, Mimicry~\cite{weissl2025mimicry} targets decision boundaries, and DeepHyperion~\cite{zohdinasab2021deephyperion} explores a feature map. Recent work benchmarks generative models for test input generation~\cite{maryam2025benchmarking} and validates generated inputs with a human-aligned learned validator~\cite{maryam2026deepnaqqal}, showing that validity and label preservation vary across generators. Riccio and Tonella~\cite{riccio2023invalid} study when and why test generators for DL produce invalid inputs, and Dola et al.~\cite{dola2021distribution} use a generative model to discard generated inputs that lie outside the training distribution. These approaches target unimodal classifiers. \tool\ extends seed-centric generation to image--text inputs, whose expected answer depends on both modalities, and confirms that answer with an oracle for every reported failure.

\noindent\textbf{Evaluation of VLMs.}
Benchmarks remain the main way to evaluate VLMs. Chen et al.~\cite{chen2023benchmarking} measure robustness under visual and textual corruptions, VHELM~\cite{lee2024vhelm} evaluates VLMs along multiple aspects, NaturalBench~\cite{li2024naturalbench} collects natural adversarial samples that cannot be solved from one modality alone, Rahmanzadehgervi et al.~\cite{rahmanzadehgervi2024blind} show that VLMs fail at simple visual tasks, and compositional benchmarks probe attribute binding and word order~\cite{thrush2022winoground,hsieh2023sugarcrepe}. MultiMon~\cite{tong2023mass} mines failure patterns from inputs on which a text encoder and a VLM erroneously agree. These works evaluate models on fixed or curated inputs. \tool\ complements them by generating new inputs from each seed.

\noindent\textbf{Metamorphic and Search-Based Testing of Multimodal Systems.}
Metamorphic testing~\cite{chen2018metamorphic,segura2016survey} checks relations between the outputs of related inputs when the expected output is unknown. MetaVQA~\cite{yuan2021perception} synthesizes counting and existence questions from detected objects and checks relations under question transformations, such as splitting or negating a question, and under image transformations, such as inserting or removing objects; each relation transforms one modality. MetaRA~\cite{xu2026metara} applies metamorphic relations to the image--question inputs of MLLM-based VQA systems and reports the failure patterns they reveal. For image captioning, MetaIC~\cite{yu2022metaic} inserts objects into images and ROME~\cite{yu2023rome} removes them, and both check how the captions of the transformed images should relate. Metamorphic relations have also been defined for vision--language--action models in robotic manipulation~\cite{mt4vla2026}, and search-based testing has been applied to VLMs for in-car scene understanding~\cite{sorokin2026search}. \tool\ also builds on an invariance relation. It differs in that it checks the precondition of the relation, namely that the expected answer is preserved, for every reported failure, and it pairs each joint failure with matched single-modality controls whose answers are confirmed.
We compare with the relations of MetaRA that apply to our setting (Section~\ref{sec:baselines}).

\noindent\textbf{Adversarial Attacks on VLMs.}
Adversarial attacks optimize perturbations to induce mispredictions under a threat model~\cite{zhao2023evaluating,schlarmann2024robust}. Co-Attack~\cite{zhang2022towards} perturbs the text first and then the image, guided by the adversarial text, and SGA~\cite{lu2023set} alternates guidance between the two modalities to improve transferability. VLATTACK~\cite{yin2023vlattack} uses a pretrained model to attack black-box models fine-tuned from it, with single-modality image and text attacks followed by an iterative cross-search over both modalities. Attacks maximize attack success under a perturbation bound. \tool\ pursues a different goal: it generates controlled test inputs and reports failures with oracle-confirmed expected answers and attribution evidence.

\noindent\textbf{Text Perturbation and NLP Testing.}
Text adversarial methods such as DeepWordBug~\cite{gao2018black}, SCPN~\cite{iyyer2018adversarial}, PWWS~\cite{ren2019generating}, and TextFooler~\cite{jin2020bert}, and behavioral test suites such as CheckList~\cite{ribeiro2020checklist}, perturb text for robustness and capability testing. They supply the lexical operators that our text pipeline builds on, and TextFooler is the text-side baseline in our evaluation.

\noindent\textbf{Combinatorial Interaction Testing.}
Combinatorial interaction testing (CIT)~\cite{kuhn2004software,nie2011survey} builds on the observation that many faults are triggered by interactions among a few input factors. A joint failure is such an interaction between two factors, the image and the text. In classical CIT, factors have small predefined domains and covering arrays enumerate their combinations. In \tool, the levels of each factor are generated variants, so \tool\ selects which combinations to test and attributes each failure with four-cell evidence.

\begin{table*}[t]
\centering
\footnotesize
\caption{Comparison of \tool\ with representative approaches (I = image, T = text). In the last column, n/a means that the approach does not test joint inputs, and no means that it does, without matched single-modality controls.}
\label{tab:paradigms}
\renewcommand{\arraystretch}{0.82}
\setlength{\tabcolsep}{4pt}
\begin{tabular*}{\textwidth}{@{\extracolsep{\fill}}L{2.6cm} L{2.9cm} L{1.3cm} L{3.2cm} L{2.2cm}}
\toprule
\textbf{Approach} & \textbf{Test inputs} & \textbf{Mutated} & \textbf{Expected answer of generated inputs} & \textbf{Matched four-cell confirmation} \\
\midrule
Chen et al.~\cite{chen2023benchmarking} & predefined corruptions of dataset inputs & I or T & assumed unchanged & n/a \\
NaturalBench~\cite{li2024naturalbench} & curated natural adversarial pairs & -- & human-annotated & n/a \\
MultiMon~\cite{tong2023mass} & mined from existing corpora & -- & summarized post hoc & n/a \\
MetaVQA~\cite{yuan2021perception} & relation-driven transformations of one modality & I or T & defined by the relation & n/a \\
MetaRA~\cite{xu2026metara} & relation-driven transformations & I and T & defined by the relation & no \\
Co-Attack~\cite{zhang2022towards}, SGA~\cite{lu2023set} & white-box optimization with cross-modal guidance & I+T & assumed under perturbation bounds & no \\
VLATTACK~\cite{yin2023vlattack} & per-example optimization with cross-search & I, T, I+T & assumed under perturbation bounds & no \\
SINVAD~\cite{kang2020sinvad}, CIT4DNN~\cite{dola2024cit4dnn} & latent-space generation & I (unimodal models) & not checked & n/a \\
DeepNaqqal~\cite{maryam2026deepnaqqal} & validator for generated inputs & I (unimodal models) & not addressed; input validity predicted by a learned validator & n/a \\
\midrule
\textbf{\tool} & seed-centric generation; single-modality tests, then joint tests & I, T, I+T & filtered, then confirmed by an automated oracle & yes \\
\bottomrule
\end{tabular*}
\end{table*}

\section{Problem Formulation}
\label{sec:problem}

\subsection{Inputs and Expected Answers}
A VLM $M$ maps an image--text pair $s=(x,t)$ to an output $M(s)$. A task-specific normalization $\norm(\cdot)$ maps outputs to comparable answers: binary labels for ITM and VR, and canonical short answers for VQA following the official evaluation script~\cite{antol2015vqa,vqaEvaluation}. A \emph{seed} $s_0=(x_0,t_0)$ is a dataset input. A VR input also contains a second image, which stays fixed in all inputs derived from a seed and is omitted from the notation.

The \emph{reference answers} $A_0$ of a seed come from the dataset: the label for ITM and VR, and for VQA the answers that receive full credit under the official metric, that is, answers given by at least four of the ten annotators. The \emph{expected answer} $A(s)$ of an input $s$ is the set of answers that are correct for $s$. It can contain more than $A_0$, such as a synonym of a reference answer, and for a generated input it is unknown until the oracle establishes it. $M$ is \emph{correct} on $s$ if $\norm(M(s))\in A(s)$, and the output of $s$ is \emph{compatible} if $\norm(M(s))\in A_0$. A compatible output is correct whenever $A(s)=A(s_0)$, and an incompatible output can still be correct.

\subsection{Test Oracle}
\tool\ uses the invariance relation
\begin{equation}
A(s)=A(s_0)\ \Rightarrow\ \norm(M(s))\in A(s_0) .
\label{eq:mr}
\end{equation}
Its precondition depends on how the image and the text relate and cannot be read off the mutation. \tool\ therefore separates the check done during testing, which compares outputs only with the reference answers, from the confirmation done by an automated oracle (Section~\ref{sec:verification}), which establishes the expected answer.

\begin{definition}[Candidate]
Let $s_0$ be an admitted seed (Section~\ref{sec:reference}). A generated input $s$ is a \emph{candidate} if its output is not compatible, $\norm(M(s))\notin A_0$.
\end{definition}

\begin{definition}[Confirmed failure]
A candidate $s$ is a \emph{confirmed failure} if the oracle establishes that $s$ is a valid input with $A(s)=A(s_0)$ and that $\norm(M(s))\notin A(s)$.
\end{definition}

\begin{definition}[Attributed joint failure]
A joint input $(x,t)$ of an admitted seed $s_0$ is an \emph{attributed joint failure} if $(x,t)$ is a confirmed failure and the oracle establishes that both single-modality cells $(x,t_0)$ and $(\hat x_0,t)$ are valid inputs with expected answer $A(s_0)$ on which $M$ is correct.
\end{definition}

A candidate need not be a failure: its expected answer may have changed, the input may be invalid, or its output may be a correct answer outside $A_0$. Two single-modality variants that each preserve the answer do not guarantee that their combination does, so the oracle checks the joint cell on its own. A joint input that is a confirmed failure is not necessarily an attributed joint failure: the latter additionally requires both single-modality controls, which attribute the failure to the combination of the two mutations. We refer to the former as a joint input confirmed on its own.

\subsection{Reference Input}
\label{sec:reference}
Every visual variant passes through the VQGAN encoder and decoder, and reconstruction alone can change details such as colors, counts, or text in the image. \tool\ therefore builds all cells on the reconstructed seed $\hat x_0=D(E(x_0))$: the seed cell is $(\hat x_0,t_0)$, a text-only input is $(\hat x_0,t)$, and an image-only or joint input contains a decoded image variant. Each cell then differs from the seed cell only by the mutations it contains. A seed is \emph{admitted} for a model if the oracle confirms that reconstruction keeps its expected answer and the output of the model on $(\hat x_0,t_0)$ is compatible. Each model is also called once on the raw seed $(x_0,t_0)$, so that we can report how often reconstruction changes its output; this call is not a test.

\subsection{Interaction and Additive Failures}
\label{sec:interaction}
An attributed joint failure shows that neither mutation causes the failure alone. The two effects may still simply add up: each mutation moves the model toward a wrong answer, and only together do they cross the decision boundary. For models that expose output scores, we separate this case from an interaction between the mutations. Let $\hat y$ be the answer of $M$ on the joint input $(x,t)$, and let $\eta(s)=\max_{y\in A_0}\log p_M(y\mid s)-\log p_M(\hat y\mid s)$ be the margin of the best reference answer over $\hat y$ on a cell $s$. Adding the effects of the two mutations predicts the margin of the joint input as
\begin{equation}
\tilde\eta(x,t)=\eta(x,t_0)+\eta(\hat x_0,t)-\eta(\hat x_0,t_0).
\label{eq:additive}
\end{equation}
An attributed joint failure is an \emph{interaction failure} if $\tilde\eta(x,t)>0$, that is, if the additive prediction still prefers a reference answer, and an \emph{additive failure} otherwise. The difference $\Delta_{\mathrm{int}}=\eta(x,t)-\eta(x,t_0)-\eta(\hat x_0,t)+\eta(\hat x_0,t_0)$ is the interaction contrast of the $2\times2$ design that the four cells form. Hessel and Lee~\cite{hessel2020does} use a related additive projection to test whether a model uses cross-modal interactions on a dataset; we apply the contrast to each attributed joint failure. The analysis uses the scores only for evaluation; \tool\ itself uses normalized outputs.

\section{Approach}
\label{sec:approach}

\begin{figure*}[t!]
    \centering
    \includegraphics[width=1\linewidth]{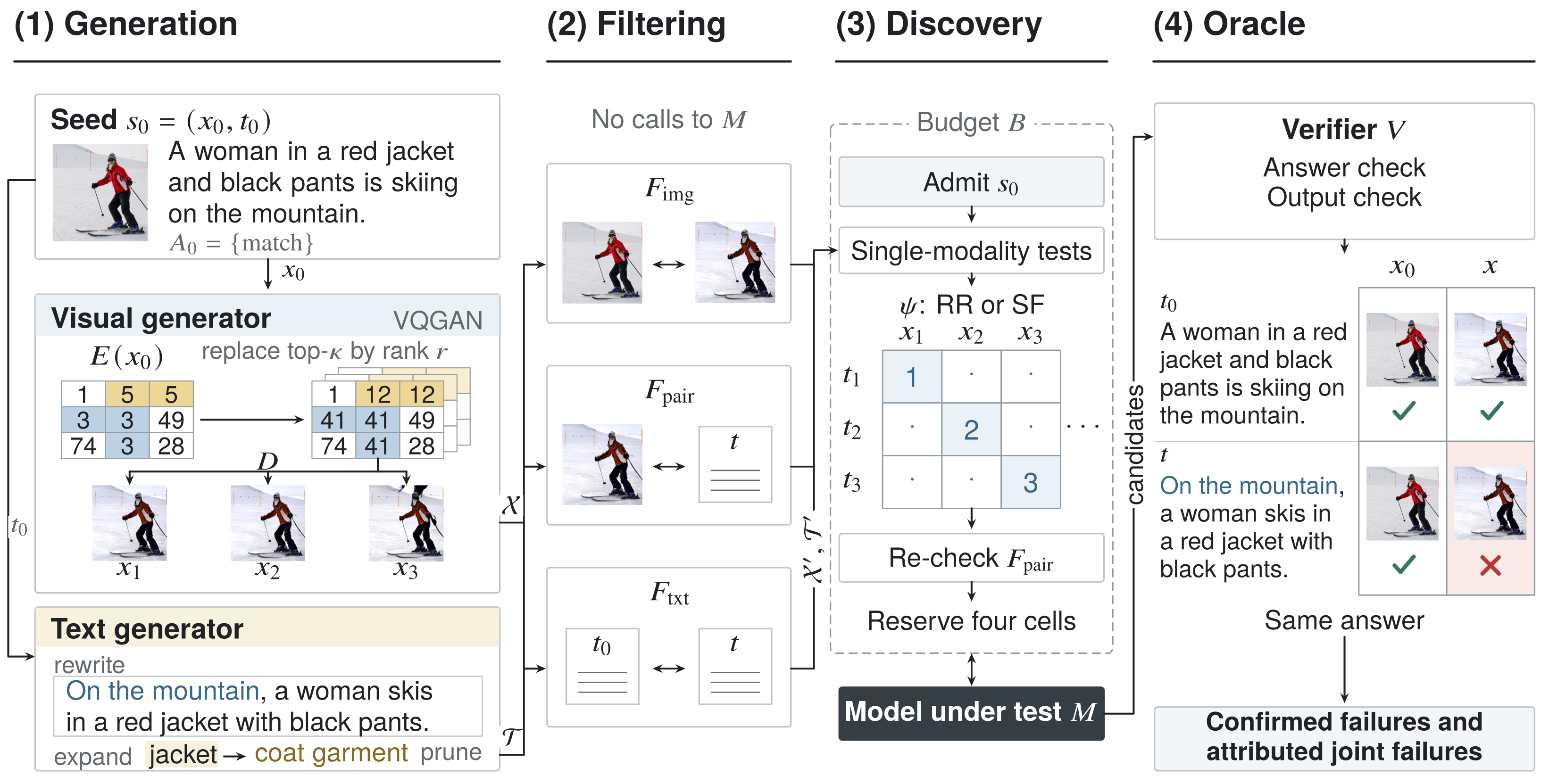}
    \vspace{-4mm}
    \caption{Overview of \tool.}
    \label{fig:overview}
    \vspace{-3mm}
\end{figure*}

\tool\ treats the model under test as a black box and accesses only its inputs and normalized outputs. Its generators and filters use pretrained, off-the-shelf models. Figure~\ref{fig:overview} gives an overview of its four stages. First, the visual and textual generators (Sections~\ref{sec:visual} and~\ref{sec:text}) emit deduplicated variants of the seed in fixed orders. Second, the relation filters (Section~\ref{sec:filters}) remove variants that are likely to change the expected answer, before any call to the model under test. Third, two-phase testing (Section~\ref{sec:discovery} and Algorithm~\ref{alg:discovery}) admits the seed, runs the single-modality tests, and then runs joint inputs in the order of a pairing strategy (Section~\ref{sec:pairing}). Fourth, the automated oracle (Section~\ref{sec:verification}) turns the candidates into confirmed failures and attributed joint failures.

\subsection{Visual Variant Generation}
\label{sec:visual}

\tool\ produces visual variants by editing the discrete latent grid of a pretrained VQGAN~\cite{esser2021taming}. The encoder maps a seed image $x_0$ to an index grid $Z_0=E(x_0)\in\{1,\dots,K\}^{h\times w}$, each index refers to a codebook entry with embedding $\mathbf{e}(c)\in\mathbb{R}^m$, and the decoder $D$ maps a grid back to an image. Intuitively, the grid describes the image as $h\times w$ cells, each labeled with one of $K$ reusable ``visual words''. Replacing a label at all cells where it occurs edits the corresponding visual content throughout the image. Because the decoder has a receptive field larger than one cell, an edit also influences neighboring content.

\noindent\textit{Index ranking.} For an index $c$ in $Z_0$, let $\mathrm{freq}_{x_0}(c)=|\{(u,v)\mid Z_0[u,v]=c\}|$. \tool\ sorts the indices by decreasing frequency, breaking ties by increasing index value, and keeps the top $\kappa$ as $\mathcal{C}_{x_0}=\langle c_1,\dots,c_\kappa\rangle$. Frequent indices cover larger regions and correspond to dominant visual prototypes, so replacing them produces observable edits with few variants.

\noindent\textit{Replacement along embedding distance.} For each $c_i\in\mathcal{C}_{x_0}$, the other codebook indices are sorted by embedding distance, $\mathcal{N}(c_i)=\langle c_i^{(1)},\dots,c_i^{(K-1)}\rangle$, with $\|\mathbf{e}(c_i)-\mathbf{e}(c_i^{(r)})\|_2$ non-decreasing in $r$. The variant at rank $r$ replaces every occurrence of each $c_i$ by $c_i^{(r)}$, for all $i$ at once, and decodes the edited grid.

\noindent\textit{Mutation strength and order.} The rank range $\{1,\dots,R\}$, with $R\le K-1$, controls mutation strength, and the number of requested variants controls how many ranks are used. \tool\ visits ranks in a fixed bit-reversal (van der Corput) order: the $j$-th variant uses rank $\sigma_R(j)$, where $\sigma_R$ enumerates the bit-reversed integers $0,\dots,2^{\lceil\log_2 R\rceil}-1$, shifted by one, and skips values above $R$. Every prefix of this order spreads nearly uniformly over the range, so requesting more variants only appends to the sequence. \tool\ removes duplicates by grid hash and, after the model's own preprocessing, by a hash of the model input; for the API models, whose preprocessing is not observable, the second hash is taken over the request sent to the API.

\subsection{Textual Variant Generation}
\label{sec:text}

\tool\ produces textual variants with a three-stage pipeline that bounds the neighborhood of a seed text $t_0$.
In the \emph{rewrite} stage, a pretrained paraphraser (PEGASUS~\cite{zhang2020pegasus}) generates $\gamma$ sentence-level rewrites.
In the \emph{expand} stage, spaCy~\cite{honnibal2020spacy} identifies the content tokens of each rewrite, excluding named entities, numbers, and punctuation, and each eligible token is replaced by one of the top-$\tau$ fill-mask predictions of RoBERTa~\cite{liu2019roberta} with the same part of speech; at most $\delta$ tokens are substituted per rewrite.
In the \emph{prune} stage, the candidates of each rewrite group are ranked by Sentence-BERT~\cite{reimers2019sbert} cosine distance to $t_0$, and the closest $\beta/\gamma$ per group are kept after exact-string deduplication. The output interleaves the groups by within-group rank, so that, as on the visual side, a larger pool only appends to a smaller one.

\subsection{Relation Filters}
\label{sec:filters}

The generators control how far a variant moves from the seed, but not whether its expected answer changes. \tool\ applies three filters before any model call. RQ1 evaluates their precision and recall against the oracle.

\paragraph{Text filter $F_{\mathrm{txt}}$.} A variant $t$ is kept if a natural language inference model~\cite{williams2018broad} predicts entailment in both directions, $\min\{P(t_0\Rightarrow t),\,P(t\Rightarrow t_0)\}\ge\theta_{\mathrm{nli}}$. Bidirectional entailment targets equivalence of meaning, which embedding distance alone does not capture.

\paragraph{Image filter $F_{\mathrm{img}}$.} A visual variant $x$ is kept if its image embedding is close to that of the seed, $\mathrm{sim}(x,x_0)\ge\rho_{\mathrm{img}}$. This removes degenerate edits.

\paragraph{Pair filter $F_{\mathrm{pair}}$.} An image--text alignment score $a(x,t)$ is used with task-specific rules:
\begin{itemize}[leftmargin=*]
\item ITM, positive seed: keep $(x,t)$ if $a(x,t)\ge\rho\cdot a(\hat x_0,t_0)$, to retain pairs that are likely to still match.
\item ITM, negative seed: keep $(x,t)$ if $a(x,t)\le a(\hat x_0,t_0)+\mu$, to discard pairs that drift toward a match.
\item VR, true statement: keep $(x,t)$ if $a(x,t)\ge\rho\cdot a(\hat x_0,t_0)$ on the mutated image.
\end{itemize}
For false VR statements and VQA questions, image--text similarity does not reflect the expected answer, so \tool\ applies only $F_{\mathrm{txt}}$ and $F_{\mathrm{img}}$ and leaves the joint relation to the oracle. Both image-based filters use OpenCLIP ViT-bigG/14~\cite{cherti2023reproducible}. We chose it over SigLIP~\cite{zhai2023siglip} because PaliGemma and Qwen3-VL build on SigLIP encoders~\cite{qwen2025qwen3vl}, and the filters should not share a vision encoder with the models under test. The thresholds are $\theta_{\mathrm{nli}}=0.80$, $\rho_{\mathrm{img}}=0.90$, $\rho=0.95$, and $\mu=0.05$, fixed on a pilot (Section~\ref{sec:budget}).

\subsection{Two-Phase Testing}
\label{sec:discovery}

\begin{algorithm}[t]
\caption{Two-phase testing for one seed and one model}
\footnotesize
\label{alg:discovery}
\begin{algorithmic}[1]
\Require seed $s_0=(x_0,t_0)$ with reference answers $A_0$ and reconstruction $\hat x_0$; model $M$; ordered sequences $\mathcal{X},\mathcal{T}$; pool size $n$; number of joint tests $n_J$; filter flag $\phi$; pairing strategy $\psi$
\Ensure call log $L$; candidates $\mathcal{F}_V,\mathcal{F}_T$; joint candidates $\mathcal{F}_J$ with four cells
\State $\mathcal{F}_V,\mathcal{F}_T,\mathcal{F}_J\gets\emptyset$ \Comment{$\call$ reuses a logged output or issues one new call}
\State $\call(x_0,t_0)$ \Comment{raw seed, not a test}
\If{$A(\hat x_0,t_0)\neq A(s_0)$ \textbf{or} $\norm(\call(\hat x_0,t_0))\notin A_0$} \State \Return seed not admitted for $M$ \EndIf
\If{$\phi$}
  \State $\mathcal{X}\gets[x\in\mathcal{X}\mid F_{\mathrm{img}}(x)\wedge F_{\mathrm{pair}}(x,t_0)]$
  \State $\mathcal{T}\gets[t\in\mathcal{T}\mid F_{\mathrm{txt}}(t)\wedge F_{\mathrm{pair}}(\hat x_0,t)]$
\EndIf
\State $\mathcal{X}'\gets$ first $n$ of $\mathcal{X}$;\quad $\mathcal{T}'\gets$ first $n$ of $\mathcal{T}$
\ForAll{$x\in\mathcal{X}'$} \Comment{single-modality phase}
  \State $\textsc{Test}(x,t_0)$
\EndFor
\ForAll{$t\in\mathcal{T}'$}
  \State $\textsc{Test}(\hat x_0,t)$
\EndFor
\State $C_J\gets 0$
\ForAll{$(x,t)\in\psi.\textsc{Order}(\mathcal{X}',\mathcal{T}',L)$ while $C_J<n_J$} \Comment{joint phase}
  \If{$\phi\wedge\neg F_{\mathrm{pair}}(x,t)$} \textbf{continue} \Comment{joint re-check, no call}
  \EndIf
  \State $C_J\gets C_J+1$
  \If{$\textsc{Test}(x,t)$}
    \State add $\langle(\hat x_0,t_0),(x,t_0),(\hat x_0,t),(x,t)\rangle$ and the four outputs to $\mathcal{F}_J$
  \EndIf
\EndFor
\State \Return $L,\mathcal{F}_V,\mathcal{F}_T,\mathcal{F}_J$
\end{algorithmic}
\end{algorithm}

Algorithm~\ref{alg:discovery} runs the tests for one seed. $\textsc{Test}(u)$ obtains the output of $u$ through $\call$, logs $u$ as a test, adds $u$ to $\mathcal{F}_V$ or $\mathcal{F}_T$ if it is a single-modality candidate, and returns whether $u$ is a candidate. The strategies $\psi\in\{\textsc{RR},\textsc{SF}\}$ are those of \tool\ (Section~\ref{sec:pairing}); the other values of $\psi$ and the flag $\phi$ implement the joint baselines of Section~\ref{sec:baselines} within the same procedure.

Admission (Section~\ref{sec:reference}) is per model, so each model is tested on the inputs it answers with a reference answer.
The single-modality phase tests the first $n$ variants of each filtered sequence, and the joint phase combines exactly these variants. Because every single-modality input is tested before the joint phase, each joint candidate has complete four-cell evidence when it is found. The single-modality phase also yields the single-modality failures of the run.
The pair filter is applied again to the joint input, because two variants that pass it separately need not pass it together. A combination that fails the re-check is skipped without a call.

\subsection{Pairing Strategies}
\label{sec:pairing}
\tool\ provides two strategies for ordering the pairs in $\mathcal{X}'\times\mathcal{T}'$.

\paragraph{Round-robin (RR)} Let $P$ be the longer of the two pools and $Q$ the shorter one, with sizes $n_P\ge n_Q$. The $k$-th pair, $k=0,1,\dots$, combines $P[k \bmod n_P]$ with $Q[((k \bmod n_P)+\lfloor k/n_P\rfloor) \bmod n_Q]$. Each round of $n_P$ pairs uses every element of the longer pool once and shifts the shorter pool by one position, so after $n_Q$ rounds every pair has appeared exactly once.

\paragraph{Stable-first (SF)} SF uses the results of the single-modality tests as feedback. Pairs whose image-only and text-only inputs both have compatible outputs come first, in RR order, followed by the remaining pairs in RR order. Because an attributed joint failure requires correct single-modality inputs, SF first tests the combinations whose parts single-modality testing has not broken.

\subsection{Oracle}
\label{sec:verification}
\tool\ confirms candidates with an automated oracle built on a VLM judge $V$. The judge receives the seed and its reference answers $A_0$ as a reference, so it decides whether an input keeps a known answer instead of answering the input from scratch. For each generated cell $s$, the oracle runs two checks in separate calls, whether or not the output of the cell is compatible.
$V$ receives the seed, $A_0$, and $s$, and returns \emph{preserved} if $s$ is a valid input with the expected answer of the seed, \emph{changed} if its expected answer differs, \emph{invalid} if $s$ is not a meaningful input for the task, or \emph{abstain}. It does not see any model output, so the output cannot influence the expected answer.
$V$ then receives the seed, $A_0$, $s$, and the output, and returns \emph{correct} if the output is a correct answer for $s$, such as a synonym of a reference answer, \emph{incorrect}, or \emph{abstain}. A cell is correct only if the first check returns preserved and the second returns correct. The two verdicts can contradict each other when the output is compatible: preserved means that the reference answers remain correct, and a compatible output is one of them. The oracle keeps the verdict of the output check in this case, so such a cell counts as incorrect and rejects the joint candidate that contains it (Section~\ref{sec:rq3}).
A candidate is a confirmed failure if its answer check returns preserved and its output check returns incorrect. A joint candidate is an attributed joint failure if the answer check returns preserved for its three generated cells and the output check returns correct for both single-modality inputs and incorrect for the joint input. Abstentions do not confirm failures. The judge is not told which method produced an input. The oracle also runs the answer check on the reconstructed seed of every sampled seed.
On API models, a failure must also be reproducible. The seed cell $(\hat x_0,t_0)$ and each cell of a candidate are called five more times, and a cell is correct or incorrect if at least three of the five new outputs are. A confirmed failure then requires a correct seed and an incorrect candidate, and an attributed joint failure also requires correct single-modality inputs. These calls are repeat calls $C_{\mathrm{rep}}$.

\section{Evaluation Design}
\label{sec:design}

\subsection{Research Questions}
\begin{itemize}[leftmargin=*,noitemsep,topsep=1pt]
\item \textbf{RQ1 (Validity):}
How many candidates does the oracle confirm as failures, and how precise is it?

\item \textbf{RQ2 (Testing utility and cost):}
How effectively and efficiently does \tool\ find single-modality failures and attributed joint failures?

\item \textbf{RQ3 (Joint failures):}
What attributed joint failures does \tool\ reveal?

\item \textbf{RQ4 (Sensitivity and generalization):}
How do the results vary across design choices and VLMs?
\end{itemize}

RQ1 breaks the candidates down by oracle outcome, measures the precision of the oracle and of its rejections against human labels, and evaluates the filters and VQGAN reconstruction.
RQ2 compares the single-modality failures with two single-modality baselines and the attributed joint failures with five joint baselines, under the same budget.
RQ3 breaks the joint candidates down by four-cell outcome, separates interaction from additive failures, and categorizes the failures.
RQ4 varies the design choices of \tool\ and repeats the joint comparison on four general-purpose VLMs.

\subsection{Datasets and Models}
\label{sec:seeds}
Following~\cite{chen2023benchmarking}, we use ITM on MSCOCO~\cite{lin2014microsoft}, VR on NLVR2~\cite{suhr2019corpus}, and VQA on VQA~v2~\cite{goyal2017making}; each dataset input is a seed. An NLVR2 input has two images and a statement; \tool\ mutates the first image, keeps the second, and gives both to the oracle. Each task has two fine-tuned models: BLIP~\cite{li2022blip} and ViLT~\cite{kim2021vilt} for ITM, BEiT-3~\cite{wang2023image} and ViLT for VR, and PaliGemma~\cite{beyer2024paligemma} and BEiT-3 for VQA.
Seeds are sampled at random from the validation splits, one per source image (per image pair for NLVR2). For ITM, half of the seeds are positive pairs and half are negative pairs. A negative pair combines an image with the caption of a random image that shares no MSCOCO object category with it, and is kept only if the judge, shown the pair without the label, judges it as not matching. Sampling continues until each task--model pair has 100 admitted seeds.

\subsection{Baselines}
\label{sec:baselines}
Table~\ref{tab:baselines} lists the compared methods. All methods have the same budget (Section~\ref{sec:budget}), and all except RND and MetaRA-c apply the relation filters.

\noindent\textit{Single-modality methods.} SINVAD-VQ adapts SINVAD~\cite{kang2020sinvad} from a VAE to the VQGAN. Its genetic search mutates the continuous encoder output before quantization and minimizes the model's score for the seed answer. A variant rejected by the filters receives fitness $+\infty$ without a call, and every evaluated image is a test. TextFooler~\cite{jin2020bert} ranks the words of $t_0$ by how much deleting them lowers the score for the seed answer and then tries synonym substitutions word by word; every tried substitution is a test. It does not stop at the first success but continues until all words are processed or the budget is spent. Both baselines use output scores, which \tool\ does not: the matching score for ITM, the class probability for VR, and the log-probability of the seed answer for VQA, where generating an answer and scoring the seed answer are separate calls. \tool-V and \tool-T test the first $B$ filtered image or text variants of our generators.

\noindent\textit{Joint methods.} All joint methods run Algorithm~\ref{alg:discovery} with the same $n$ and $n_J$. RND pairs unfiltered pools of our generators in random order. CRND pairs the filtered pools of \tool\ in random order. SR first pairs, in random order, the variants whose single-modality inputs have compatible outputs, so it differs from \tool\ (SF) only in the order within these pairs. EXT runs SINVAD-VQ and TextFooler for $n$ calls each as its single-modality phase, uses the variants they test as its pools, and pairs them in the stable-first order of \tool; it differs from \tool\ (SF) only in the variants it combines. MetaRA-c applies the relations of MetaRA~\cite{xu2026metara} that can be instantiated as seed-relative transformations on VQA: its single-modality phase tests the image and question transformations, and its joint phase combines them. We exclude CIT4DNN~\cite{dola2024cit4dnn} and Mimicry~\cite{weissl2025mimicry}, which require a low-dimensional or class-conditional latent. We also do not compare with multimodal adversarial attacks such as Co-Attack~\cite{zhang2022towards}, SGA~\cite{lu2023set}, and VLATTACK~\cite{yin2023vlattack}. On the image side, they optimize pixel-level perturbations against a surrogate model to maximize attack success under a perturbation bound, and they assume rather than check that the expected answer is preserved. Their confirmed failures would therefore measure attack strength rather than the ability to generate semantically varied test inputs. On the text side, the word-level substitutions such attacks use are represented by TextFooler.

\begin{table*}[t]
\centering
\footnotesize
\caption{Compared methods. $\phi$: relation filters applied. The first four methods test one modality only.}
\label{tab:baselines}
\renewcommand{\arraystretch}{0.82}
\setlength{\tabcolsep}{3.5pt}
\begin{tabular*}{\textwidth}{@{\extracolsep{\fill}}lllcl}
\toprule
\textbf{Method} & \textbf{Image side} & \textbf{Text side} & $\phi$ & \textbf{Joint inputs} \\
\midrule
SINVAD-VQ & latent search (scores) & -- & yes & -- \\
\textbf{\tool-V} & codebook replacement & -- & yes & -- \\
TextFooler & -- & synonym search (scores) & yes & -- \\
\textbf{\tool-T} & -- & rewrite--expand--prune & yes & -- \\
\midrule
RND & \tool-V & \tool-T & no & random \\
CRND & \tool-V & \tool-T & yes & random \\
SR & \tool-V & \tool-T & yes & random, stable first \\
EXT & SINVAD-VQ & TextFooler & yes & stable-first \\
MetaRA-c (VQA) & relation-defined & relation-defined & -- & relation-defined \\
\midrule
\textbf{\tool\ (RR)} & \tool-V & \tool-T & yes & round-robin \\
\textbf{\tool\ (SF)} & \tool-V & \tool-T & yes & stable-first \\
\bottomrule
\end{tabular*}
\end{table*}

\subsection{Metrics}
\label{sec:metrics}
The oracle judges every candidate. For each task, model, and method, we report:
\begin{itemize}[leftmargin=*,noitemsep,topsep=1pt]
\item $\cfk=1000\,N_{\cf}/(B\,|S|)$, the confirmed single-modality failures per 1,000 calls of the budget, where $N_{\cf}$ counts the confirmed failures of one modality over the admitted seeds $S$; a method that stops early is still divided by the full budget;
\item $\vjfk=1000\,N_{\vjf}/(B\,|S|)$, the attributed joint failures per 1,000 calls of the joint budget;
\item budget per AJF, $1000/\vjfk$ averaged over the task--model pairs; for a method that stops early, we also give the calls it actually makes per attributed joint failure;
\item budget used, the calls made as a share of the budget, in the single-modality or the joint phase;
\item tested stable pairs, the share of joint calls on pairs whose image-only and text-only inputs both have compatible outputs;
\item SeedCov, the share of admitted seeds with at least one confirmed failure, computed separately for single-modality failures and attributed joint failures;
\item VarCov, the mean number of distinct image and text variants per seed that take part in at least one attributed joint failure;
\item confirmation rate, the share of candidates that the oracle confirms as failures;
\item interaction share, the share of attributed joint failures that are interaction failures (Section~\ref{sec:interaction}).
\end{itemize}
For ITM and VR, $\eta$ is the difference between the logits of the two labels. For VQA, it uses the log-probabilities of the answers under teacher forcing, averaged over their tokens. For the five classification models, $\eta(x,t)\le0$ holds by construction, because $\hat y$ is their highest-scoring answer. PaliGemma generates its answers, so we report how many of its attributed joint failures have $\eta(x,t)>0$ and exclude them from the margin analysis.

\noindent\textbf{Filters and reconstruction.} We measure the precision and recall of each filter against the answer check of the oracle on 1,000 unfiltered variants per task and modality, sampled at random from the generator outputs. For reconstruction, we report the share of sampled seeds whose reconstruction changes the expected answer, which are not admitted, and the share of admitted seeds on which the model output differs between the raw and the reconstructed seed.

\subsection{Settings and Implementation}
\label{sec:budget}
\label{sec:impl}
\noindent\textbf{Budgets.} The primary experiments use a budget of $B=1{,}000$ calls to the model under test. A single-modality method makes $B$ calls on inputs of its modality, including the calls a search method makes to score its options. A joint method makes $B$ single-modality calls, $n=500$ per modality, and then $B$ joint calls, so $n_J=B$ in Algorithm~\ref{alg:discovery}. Single-modality failures are thus always counted over $B$ single-modality calls and attributed joint failures over $B$ joint calls. A method whose pool or search space is exhausted stops early.

\noindent\textbf{Cost accounting.} Every call to the model under test is logged with its run, its role, and its key. A \emph{run} applies one method to one seed and one model; a key hashes the model input and the request type. Within a run, each key is called once and its result is reused, so a single-modality input that also serves as four-cell evidence is charged once. A run is charged for every call it issues, even if another run issued the same input. Calls to the raw seed and the seed cell, the repeat calls $C_{\mathrm{rep}}$ of stochastic models, and the calls of the judge are not charged to $B$.

\noindent\textbf{Parameters.} The VQGAN is the pretrained Gumbel-quantized model of Taming Transformers~\cite{esser2021taming} with downsampling factor 8 and $K=8{,}192$ codebook entries. Images are resized to $256\times256$ pixels, which gives a $32\times32$ index grid. The visual generator uses $\kappa=15$ and the full rank range $R=8{,}191$; the text generator uses $\gamma=\tau=\delta=5$ and keeps $\beta=4{,}000$ variants. The filters use $\theta_{\mathrm{nli}}=0.80$, $\rho_{\mathrm{img}}=0.90$, $\rho=0.95$, and $\mu=0.05$. SINVAD-VQ uses the population size and genetic operators of the original implementation, and TextFooler its default synonym sets and thresholds. The stochastic methods (SINVAD-VQ, RND, CRND, SR, and EXT) run five times on the local models; outputs of deterministic models are cached across runs.

\noindent\textbf{Oracle.} The judge $V$ is Llama-4-Maverick~\cite{meta2025maverick}, from a family to which none of the models under test belongs. Its prompts and the filter thresholds were fixed on a pilot with 20 seeds per task, which is excluded from the evaluation.

\noindent\textbf{Statistics.} The unit of analysis is the source image, or the image pair for NLVR2. We compare methods by paired differences with 95\% percentile intervals from a paired cluster bootstrap over the 100 source images of a task--model pair, with 10,000 resamples~\cite{efron1994bootstrap}; for a stochastic method, the five runs are averaged per seed before resampling. Two-sided $p$-values are $\min\{1,2\min(p^-,p^+)\}$, where $p^-$ and $p^+$ are the shares of resampled differences at most and at least zero. Holm's procedure~\cite{holm1979simple} corrects them within Table~\ref{tab:rq2}, which marks the significant differences.

\noindent\textbf{Implementation.} The task-specific models use their fine-tuned checkpoints, and Qwen3-VL uses its public Hugging Face checkpoints. All local models run in bf16, with greedy decoding for generative outputs. $F_{\mathrm{txt}}$ uses \texttt{roberta-large-mnli}, and $F_{\mathrm{img}}$ and $F_{\mathrm{pair}}$ use OpenCLIP ViT-bigG/14 trained on LAION-2B. Experiments ran on Ubuntu 24.04.3 LTS with an AMD Ryzen Threadripper PRO 7985WX CPU, 502\,GiB of DDR5 memory, and two NVIDIA RTX 6000 Ada GPUs.

\section{Results}
\label{sec:results}

\subsection{RQ1: Validity}
\label{sec:rq1}

We run the ten methods of Table~\ref{tab:baselines} on the six task--model pairs and MetaRA-c on the two VQA pairs, with 100 admitted seeds per pair and $B=1{,}000$. The five stochastic methods run five times, which gives 30 runs per seed (31 on VQA) and 18,200 runs in total.

\subsubsection{Candidate Outcomes}
Depending on the pair, 61.0\% (ViLT on VR) to 84.0\% (PaliGemma) of the sampled seeds are admitted. Table~\ref{tab:rq1} breaks down the candidates of \tool\ (SF). The oracle confirms 121,235 of the 148,314 candidates (81.7\%) as failures, from 76.7\% on ViLT for VR to 84.3\% on ViLT for ITM. Of the remaining candidates, 21,961 (14.8\%) have a changed expected answer, 3,065 (2.1\%) are invalid or abstained, and 2,053 (1.4\%) have outputs that the oracle judges correct despite being outside the reference-answer set.

On every pair, image-only candidates have the highest confirmation rate (86.3\% to 91.6\%), text-only candidates are 4.8 to 10.0 percentage points lower, and joint inputs are lowest (75.0\% to 83.0\%). The joint column judges the joint input alone; Section~\ref{sec:rq3} adds the single-modality controls. The text gap is largest on VR, where text-only candidates reach 77.0\% and 77.1\%: VR statements often hinge on counts and quantifiers, and a rewrite that changes one of them changes the label. On ITM and VR the answer is binary, so an output outside $A_0$ is wrong unless the expected answer changed. On VQA, 2,053 candidates (3.2\%) have an output outside $A_0$ that is still correct, such as a synonym of a reference answer, and 6.4\% of the output changes stay within $A_0$ and are not candidates. A count of output changes reports both as failures.

\begin{table*}[t]
\centering
\footnotesize
\caption{RQ1: oracle outcomes of the candidates of \tool\ (SF), and the confirmation rate per input type.}
\label{tab:rq1}
\renewcommand{\arraystretch}{0.82}
\setlength{\tabcolsep}{2pt}
\begin{tabular*}{\textwidth}{@{\extracolsep{\fill}}llrrrrrrrr}
\toprule
 & & & & & & & \multicolumn{3}{c}{\textbf{Confirmation rate (\%)}} \\
\cmidrule{8-10}
\textbf{Task} & \textbf{Model} & \textbf{Candidates} & \makecell{\textbf{Invalid /}\\\textbf{abstain}} & \makecell{\textbf{Answer}\\\textbf{changed}} & \makecell{\textbf{Output}\\\textbf{correct}} & \makecell{\textbf{Confirmed}\\\textbf{failure}} & \makecell{\textbf{image-}\\\textbf{only}} & \makecell{\textbf{text-}\\\textbf{only}} & \makecell{\textbf{joint}\\\textbf{input}} \\
\midrule
ITM & BLIP      & 22,776 & 308 & 3,399 & 0     & 19,069 & 91.6 & 86.8 & 82.0 \\
ITM & ViLT      & 25,253 & 356 & 3,608 & 0     & 21,289 & 90.7 & 85.4 & 83.0 \\
VR  & BEiT-3    & 19,164 & 514 & 3,686 & 0     & 14,964 & 87.0 & 77.0 & 76.7 \\
VR  & ViLT      & 17,179 & 482 & 3,524 & 0     & 13,173 & 86.3 & 77.1 & 75.0 \\
VQA & PaliGemma & 33,698 & 724 & 3,953 & 1,091 & 27,930 & 89.3 & 83.2 & 81.7 \\
VQA & BEiT-3    & 30,244 & 681 & 3,791 & 962   & 24,810 & 88.8 & 82.7 & 80.7 \\
\bottomrule
\end{tabular*}
\end{table*}

\subsubsection{Oracle Precision}
For each method, we sample reported failures stratified by task, input type (image-only, text-only, joint), and verdict (confirmed failure, and for joint candidates also attributed joint failure), pooling the two models of a task. We label up to 40 failures per stratum for \tool-V, \tool-T, CRND, SR, \tool\ (RR), and \tool\ (SF), and up to 20 for the other methods. We also label 20 rejected joint candidates of \tool\ (SF) for each task and reason of rejection, 200 in total, and 30 reconstructed seeds per task.

Nine graduate students in computer science who are not authors label these inputs. Two of them label each input independently, and a third resolves disagreements; an input on which no two agree counts as rejected. The annotators follow the two checks of the oracle: they decide the expected answer before they see the model output, and they see neither the oracle verdict nor the method. Before adjudication, Krippendorff's $\alpha$ is 0.82 on the answer check and 0.88 on the output check.
Precision is the share of sampled failures that the annotators agree with, with a 95\% Clopper--Pearson interval per stratum. A confirmed-failure estimate that spans several input types weights the stratum precisions, and their interval endpoints, by the numbers of reported failures in the strata.

Table~\ref{tab:oracle} lists the precision of every method; each AJF cell rests on 20 to 39 labeled failures. Precision ranges from 85.0\% to 93.9\% for confirmed failures and from 80.0\% to 94.6\% for attributed joint failures; for \tool\ (SF), it ranges from 88.8\% to 94.6\%. The lowest value, 80.0\% for attributed joint failures on VR, belongs to RND and EXT. The oracle errors found by the annotators are missed changes of the expected answer, correct answers in other wording judged incorrect, and wrong attribution or ambiguity in joint failures. On the 90 reconstructed seeds, the answer check agrees with the annotators in 95.6\% of the cases, and the annotators agree with 85.0\% to 95.0\% of the rejections in every stratum.

\begin{table*}[t]
\centering
\footnotesize
\caption{RQ1: oracle precision in percent with 95\% intervals, estimated from samples drawn separately for each method and pooled over the two models of each task. CF: confirmed failures; AJF: attributed joint failures. CF precision is shown for the single-modality methods and \tool\ (SF).}
\label{tab:oracle}
\renewcommand{\arraystretch}{0.82}
\setlength{\tabcolsep}{1.5pt}
\begin{tabular*}{\textwidth}{@{\extracolsep{\fill}}lcccccc}
\toprule
 & \multicolumn{3}{c}{\textbf{CF}} & \multicolumn{3}{c}{\textbf{AJF}} \\
\cmidrule{2-4}\cmidrule{5-7}
\textbf{Method} & \textbf{ITM} & \textbf{VR} & \textbf{VQA} & \textbf{ITM} & \textbf{VR} & \textbf{VQA} \\
\midrule
SINVAD-VQ
& 90.0 {\scriptsize[68.3,98.8]}
& 85.0 {\scriptsize[62.1,96.8]}
& 90.0 {\scriptsize[68.3,98.8]}
& -- & -- & -- \\

\textbf{\tool-V}
& \textbf{92.5} {\scriptsize[79.6,98.4]}
& 87.5 {\scriptsize[73.2,95.8]}
& 92.5 {\scriptsize[79.6,98.4]}
& -- & -- & -- \\

TextFooler
& 85.0 {\scriptsize[62.1,96.8]}
& 85.0 {\scriptsize[62.1,96.8]}
& 90.0 {\scriptsize[68.3,98.8]}
& -- & -- & -- \\

\textbf{\tool-T}
& 90.0 {\scriptsize[76.3,97.2]}
& 87.5 {\scriptsize[73.2,95.8]}
& 92.5 {\scriptsize[79.6,98.4]}
& -- & -- & -- \\

RND
& --
& --
& --
& 85.0 {\scriptsize[62.1,96.8]}
& 80.0 {\scriptsize[56.3,94.3]}
& 85.0 {\scriptsize[62.1,96.8]} \\

CRND
& --
& --
& --
& 87.9 {\scriptsize[71.8,96.6]}
& 85.7 {\scriptsize[69.7,95.2]}
& 90.3 {\scriptsize[74.2,98.0]} \\

SR
& --
& --
& --
& 91.2 {\scriptsize[76.3,98.1]}
& 86.8 {\scriptsize[71.9,95.6]}
& 91.9 {\scriptsize[78.1,98.3]} \\

EXT
& --
& --
& --
& 85.0 {\scriptsize[62.1,96.8]}
& 80.0 {\scriptsize[56.3,94.3]}
& 85.0 {\scriptsize[62.1,96.8]} \\

MetaRA-c
& --
& --
& --
& --
& --
& 85.0 {\scriptsize[62.1,96.8]} \\

\textbf{\tool\ (RR)}
& --
& --
& --
& 91.2 {\scriptsize[76.3,98.1]}
& 89.2 {\scriptsize[74.6,97.0]}
& 91.2 {\scriptsize[76.3,98.1]} \\

\textbf{\tool\ (SF)}
& 92.0 {\scriptsize[78.8,98.2]}
& \textbf{88.8} {\scriptsize[74.5,96.6]}
& \textbf{93.9} {\scriptsize[81.0,99.0]}
& \textbf{92.3} {\scriptsize[79.1,98.4]}
& \textbf{89.5} {\scriptsize[75.2,97.1]}
& \textbf{94.6} {\scriptsize[81.8,99.3]} \\
\bottomrule
\end{tabular*}
\end{table*}

\subsubsection{Filters and Reconstruction}
Of the 6,000 unfiltered variants, 1,000 per task and modality, $F_{\mathrm{txt}}$ and $F_{\mathrm{img}}$ are evaluated on the 3,000 text and the 3,000 image variants, and $F_{\mathrm{pair}}$ on the 3,000 variants of ITM and of true VR statements, to which it applies. Before filtering, 78.8\% of the text variants and 86.5\% of the image variants preserve the expected answer. $F_{\mathrm{txt}}$, $F_{\mathrm{img}}$, and $F_{\mathrm{pair}}$ keep 70.8\%, 82.7\%, and 74.6\% of the variants, with a precision of 93.1\%, 95.4\%, and 91.7\% and a recall of 83.6\%, 91.2\%, and 86.0\% for answer preservation. Each filter thus removes 70\% to 77\% of the variants whose expected answer changes and loses 9\% to 16\% of those that keep it. The answer check rejects the changed answers that pass the filters (Table~\ref{tab:rq1}).

Reconstruction changes the expected answer of 2\% to 6\% of the sampled seeds, which are not admitted. On the admitted seeds, the model output on the reconstructed seed differs from that on the raw seed for 4\% to 9\%, depending on the pair.

\subsection{RQ2: Testing Utility and Cost}
\label{sec:rq2}

\subsubsection{Single-Modality Failures}
Table~\ref{tab:rq2} lists $\cfk$ and SeedCov for the single-modality methods. \tool-V finds 1.8 to 2.1 times as many confirmed failures as SINVAD-VQ, and \tool-T 2.3 to 2.5 times as many as TextFooler, without the output scores that both baselines use. Our generators also cover more seeds, 75.7\% and 71.0\% against 63.4\% and 56.7\%. All differences are significant. SINVAD-VQ and TextFooler stop early on some seeds and use 88.9\% and 81.7\% of the budget; per call actually made, their aggregate rates rise by 12\% and 22\%, and the gap remains. For all four methods, failures are most frequent on VQA and least frequent on VR.

\begin{table*}[t]
\centering
\footnotesize
\caption{RQ2: $\cfk$ per task--model pair; SeedCov and budget used in percent, averaged over the six pairs. $^{*}$: significant difference to \tool-V or \tool-T (5\% level, Holm).}
\label{tab:rq2}
\setlength{\tabcolsep}{3pt}
\begin{tabular*}{\textwidth}{@{\extracolsep{\fill}}lrrrrrrrr}
\toprule
 & \multicolumn{2}{c}{\textbf{ITM}} & \multicolumn{2}{c}{\textbf{VR}} & \multicolumn{2}{c}{\textbf{VQA}} & & \\
\cmidrule{2-3}\cmidrule{4-5}\cmidrule{6-7}
\textbf{Method} & \textbf{BLIP} & \textbf{ViLT} & \textbf{BEiT-3} & \textbf{ViLT} & \textbf{PaliGemma} & \textbf{BEiT-3} & \textbf{SeedCov (\%)} & \makecell{\textbf{Budget}\\\textbf{used (\%)}} \\
\midrule
SINVAD-VQ & 24.1$^{*}$ & 26.8$^{*}$ & 21.6$^{*}$ & 18.9$^{*}$ & 42.7$^{*}$ & 37.5$^{*}$ & 63.4 & 88.9 \\
\textbf{\tool-V} & \textbf{50.3} & \textbf{56.4} & \textbf{42.7} & \textbf{37.2} & \textbf{78.5} & \textbf{71.4} & \textbf{75.7} & 100.0 \\
\midrule
TextFooler & 19.4$^{*}$ & 21.8$^{*}$ & 15.8$^{*}$ & 13.9$^{*}$ & 30.6$^{*}$ & 27.5$^{*}$ & 56.7 & 81.7 \\
\textbf{\tool-T} & \textbf{46.2} & \textbf{50.1} & \textbf{38.4} & \textbf{34.1} & \textbf{74.3} & \textbf{68.0} & \textbf{71.0} & 100.0 \\
\bottomrule
\end{tabular*}
\end{table*}

\subsubsection{Attributed Joint Failures}
Table~\ref{tab:rq3} reports $\vjfk$ for the joint methods. \tool\ (SF) finds 91.4 (ViLT on VR) to 196.8 (PaliGemma) attributed joint failures per 1,000 joint calls, one per 7.6 joint calls on average, and finds at least one on 98.2\% of the seeds. All joint methods except MetaRA-c use the full joint budget.

The joint methods differ one component at a time. Applying the relation filters substantially improves the yield: CRND and \tool\ (RR) find 1.6 to 1.9 times as many attributed joint failures as RND.
Stable-first ordering raises the share of joint calls on stable pairs from 85.2\% (CRND and \tool\ (RR)) to 100\% (SR and \tool\ (SF)). It increases $\vjfk$ by 12\% to 24\% from CRND to SR and by 14\% to 27\% from \tool\ (RR) to \tool\ (SF) across the six task--model pairs; averaged over the pairs, $\vjfk$ rises from 118.9 to 137.6 and from 121.9 to 142.3, and the budget per attributed joint failure falls from 9.1 to 7.8 and from 8.8 to 7.6 calls. This gain follows from the definition of an attributed joint failure: a pair with an incompatible single-modality output cannot yield one on ITM and VR, and rarely on VQA. If every stable pair were equally likely to yield an attributed joint failure, moving the remaining 14.8\% of the joint calls to stable pairs would raise $\vjfk$ by about 17\% ($1/0.852$), close to the observed average gains of 16\% and 17\%. Stable-first ordering therefore does not make a stable pair more likely to fail; it spends the joint budget only on pairs that can yield an attributed joint failure, at no additional cost.

Within the same priority class, round-robin ordering changes $\vjfk$ by only 1\% to 5\% (\tool\ (RR) versus CRND, and \tool\ (SF) versus SR). It combines every variant of the pools equally often, whereas random ordering leaves some variants untested within the joint budget. The attributed joint failures of round-robin ordering involve more distinct variants (VarCov 97.4 / 95.1 against 83.1 / 81.3 for CRND, and 105.6 / 102.2 against 84.6 / 82.2 for SR) and cover more seeds (SeedCov 94.8\% against 92.7\%, and 98.2\% against 93.7\%).
EXT applies the same stable-first pairing strategy to variants produced by SINVAD-VQ and TextFooler. \tool\ (SF) finds 1.26 to 1.33 times as many attributed joint failures as EXT across the six task--model pairs, while its failures involve more than three times as many distinct variants per seed (VarCov 105.6 / 102.2 versus 32.1 / 29.8).
On the two VQA pairs, \tool\ (SF) finds 2.8 times as many attributed joint failures as MetaRA-c under the full-budget metric. MetaRA-c exhausts the combinations of its relations after using 51.4\% of the joint budget on average. When normalized by the calls it actually makes, MetaRA-c finds about 128 attributed joint failures per 1,000 joint calls, or one per 7.8 calls, compared with about 185 per 1,000 calls, or one per 5.4 calls, for \tool\ (SF).

\begin{table*}[t]
\centering
\footnotesize
\caption{RQ2: $\vjfk$ per task--model pair; budget used, budget per AJF, VarCov (image / text), SeedCov, and the share of joint calls on stable pairs, averaged over the pairs a method runs on.}
\label{tab:rq3}
\renewcommand{\arraystretch}{0.82}
\setlength{\tabcolsep}{2pt}
\begin{tabular*}{\textwidth}{@{\extracolsep{\fill}}lrcccccccccc}
\toprule
 & \multicolumn{2}{c}{\textbf{ITM}} & \multicolumn{2}{c}{\textbf{VR}} & \multicolumn{2}{c}{\textbf{VQA}} & & & & & \\
\cmidrule{2-3}\cmidrule{4-5}\cmidrule{6-7}
\textbf{Method} & \textbf{BLIP} & \textbf{ViLT} & \textbf{BEiT-3} & \textbf{ViLT} & \textbf{PaliGemma} & \textbf{BEiT-3} & \makecell{\textbf{Budget}\\\textbf{used (\%)}} & \makecell{\textbf{Budget}\\\textbf{per AJF}} & \makecell{\textbf{VarCov}\\\textbf{(I / T)}} & \makecell{\textbf{SeedCov}\\\textbf{(\%)}} & \makecell{\textbf{Tested stable}\\\textbf{pairs (\%)}} \\
\midrule
RND & 65.9 & 72.3 & 48.4 & 41.7 & 101.6 & 89.4 & 100.0 & 15.8 & 34.6 / 33.1 & 84.7 & 73.5 \\
CRND & 116.8 & 117.9 & 86.1 & 75.1 & 168.4 & 149.1 & 100.0 & 9.1 & 83.1 / 81.3 & 92.7 & 85.2 \\
SR   & 133.1 & 146.0 & 99.2 & 87.8 & 192.4 & 167.1 & 100.0 & 7.8 & 84.6 / 82.2 & 93.7 & 100.0 \\
EXT & 105.3 & 118.1 & 78.6 & 68.7 & 155.9 & 137.2 & 100.0 & 9.8 & 32.1 / 29.8 & 92.9 & 99.2 \\
MetaRA-c & -- & -- & -- & -- & 69.7 & 61.9 & 51.4 & 15.3 & 18.2 / 16.3 & 90.5 & 68.9 \\
\textbf{\tool\ (RR)} & 120.5 & 119.4 & 89.9 & 77.6 & 171.3 & 152.4 & 100.0 & 8.8 & 97.4 / 95.1 & 94.8 & 85.2 \\
\textbf{\tool\ (SF)} & \textbf{137.2} & \textbf{151.6} & \textbf{103.7} & \textbf{91.4} & \textbf{196.8} & \textbf{173.2} & 100.0 & \textbf{7.6} & \textbf{105.6 / 102.2} & \textbf{98.2} & 100.0 \\
\bottomrule
\end{tabular*}
\end{table*}

\subsection{RQ3: Joint Failures}
\label{sec:rq3}

RQ3 analyzes the joint candidates of the RQ1 runs.

\subsubsection{Four-Cell Evidence}
Table~\ref{tab:fourcell} shows why the joint candidates of \tool\ (SF) do or do not become attributed joint failures. Of its 110,152 joint candidates, 85,390 (77.5\%) are attributed joint failures, and four-cell confirmation rejects 24,762 (22.5\%).

Table~\ref{tab:fourcell} attributes each rejection to the first check that fails, starting with the joint input. Checking the joint input alone rejects 19.7\% of the joint candidates: its expected answer changes, it is invalid or the judge abstains, or, on VQA, its output is a correct answer outside $A_0$. The two single-modality controls reject a further 2.8\% of joint candidates. Among all joint candidates, 3,101 (2.8\%) exhibit conflicting oracle verdicts on at least one control: the output is compatible with the reference answers and the answer check returns preserved, but the output check judges it incorrect. Such conflicting verdicts prevent a candidate from becoming an attributed joint failure. The annotators agree with 85.0\% to 90.0\% of the sampled rejections of this type. For 9,183 candidates (8.3\%), the joint input changes the expected answer while both single-modality inputs keep it; no check of a single-modality input can detect this. VR has the lowest share of attributed joint failures (72.2\%; 73.1\% on BEiT-3 and 71.2\% on ViLT): a single-modality input changes the expected answer for 11.7\% and 12.9\% of its joint candidates, against 5.8\% to 6.5\% on the other pairs.

\begin{table*}[t]
\centering
\footnotesize
\caption{RQ3: joint candidates of \tool\ (SF) by the first check that fails, starting with the joint input, in percent of the joint candidates, pooled over the two models of each task.}
\label{tab:fourcell}
\renewcommand{\arraystretch}{0.82}
\setlength{\tabcolsep}{3pt}
\begin{tabular*}{\textwidth}{@{\extracolsep{\fill}}lrrr}
\toprule
\textbf{Task} & \makecell{\textbf{Rejected by the}\\\textbf{joint input (\%)}} & \makecell{\textbf{Further rejected by}\\\textbf{the controls (\%)}} & \makecell{\textbf{Attributed joint}\\\textbf{failures (\%)}} \\
\midrule
ITM & 17.5 & 2.5 & 80.0 \\
VR  & 24.1 & 3.7 & 72.2 \\
VQA & 18.8 & 2.6 & 78.6 \\
\midrule
All & 19.7 & 2.8 & 77.5 \\
\bottomrule
\end{tabular*}
\end{table*}

\subsubsection{Interaction and Additive Failures}
We apply the margin analysis of Section~\ref{sec:interaction} to the attributed joint failures of \tool\ (SF) (Table~\ref{tab:interaction}). On PaliGemma, 622 of the 19,683 attributed joint failures (3.2\%) have $\eta(x,t)>0$ under teacher forcing, so the sequence score does not reproduce the decoded output; we exclude them. Of the remaining 84,768, 13,999 (16.5\%) are interaction failures: the additive prediction still prefers a reference answer, but the joint input fails. Their share ranges from 13.8\% on PaliGemma to 22.6\% on ViLT for VR and is highest on VR. The other 83.5\% are additive. The median $\Delta_{\mathrm{int}}$ is positive on every pair, so the observed joint margin is usually less negative than the additive prediction.

\begin{table*}[t]
\centering
\footnotesize
\caption{RQ3: attributed joint failures of \tool\ (SF) classified by the margin analysis. Interaction: share of interaction failures; $\Delta_{\mathrm{int}}$: median with interquartile range. Both are computed over the score-classifiable failures, which exclude the 622 failures of PaliGemma with $\eta(x,t)>0$ (84,768 in all).}
\label{tab:interaction}
\setlength{\tabcolsep}{4pt}
\begin{tabular*}{\textwidth}{@{\extracolsep{\fill}}llrrc}
\toprule
\textbf{Task} & \textbf{Model} & \makecell{\textbf{Attributed joint}\\\textbf{failures}} & \makecell{\textbf{Interaction}\\\textbf{(\%)}} & \makecell{\textbf{Median $\Delta_{\mathrm{int}}$}\\\textbf{[IQR]}} \\
\midrule
ITM & BLIP & 13,716 & 16.4 & 0.121 [0.051, 0.238] \\
ITM & ViLT & 15,163 & 14.7 & 0.109 [0.043, 0.219] \\
VR & BEiT-3 & 10,368 & 21.3 & 0.154 [0.062, 0.298] \\
VR & ViLT & 9,139 & 22.6 & 0.163 [0.038, 0.312] \\
VQA & PaliGemma & 19,683 & 13.8 & 0.097 [0.038, 0.194] \\
VQA & BEiT-3 & 17,321 & 15.1 & 0.113 [0.045, 0.226] \\
\midrule
\multicolumn{2}{l}{All} & 85,390 & 16.5 & 0.124 [0.049, 0.248] \\
\bottomrule
\end{tabular*}
\end{table*}

\subsubsection{Failure Types}
Two authors open-coded a separate pilot of 30 attributed joint failures and then independently assigned one primary type to each attributed joint failure in a stratified random sample of 180, 60 per task (Cohen's $\kappa=0.79$ before adjudication). Table~\ref{tab:taxonomy} lists the five types. Attribute or entity binding and spatial or relational composition account for 98 of the 180 failures. Quantifier, count, and scope errors concentrate on VR (26 of 34).

\begin{figure*}[t]
\centering
\includegraphics[width=\textwidth]{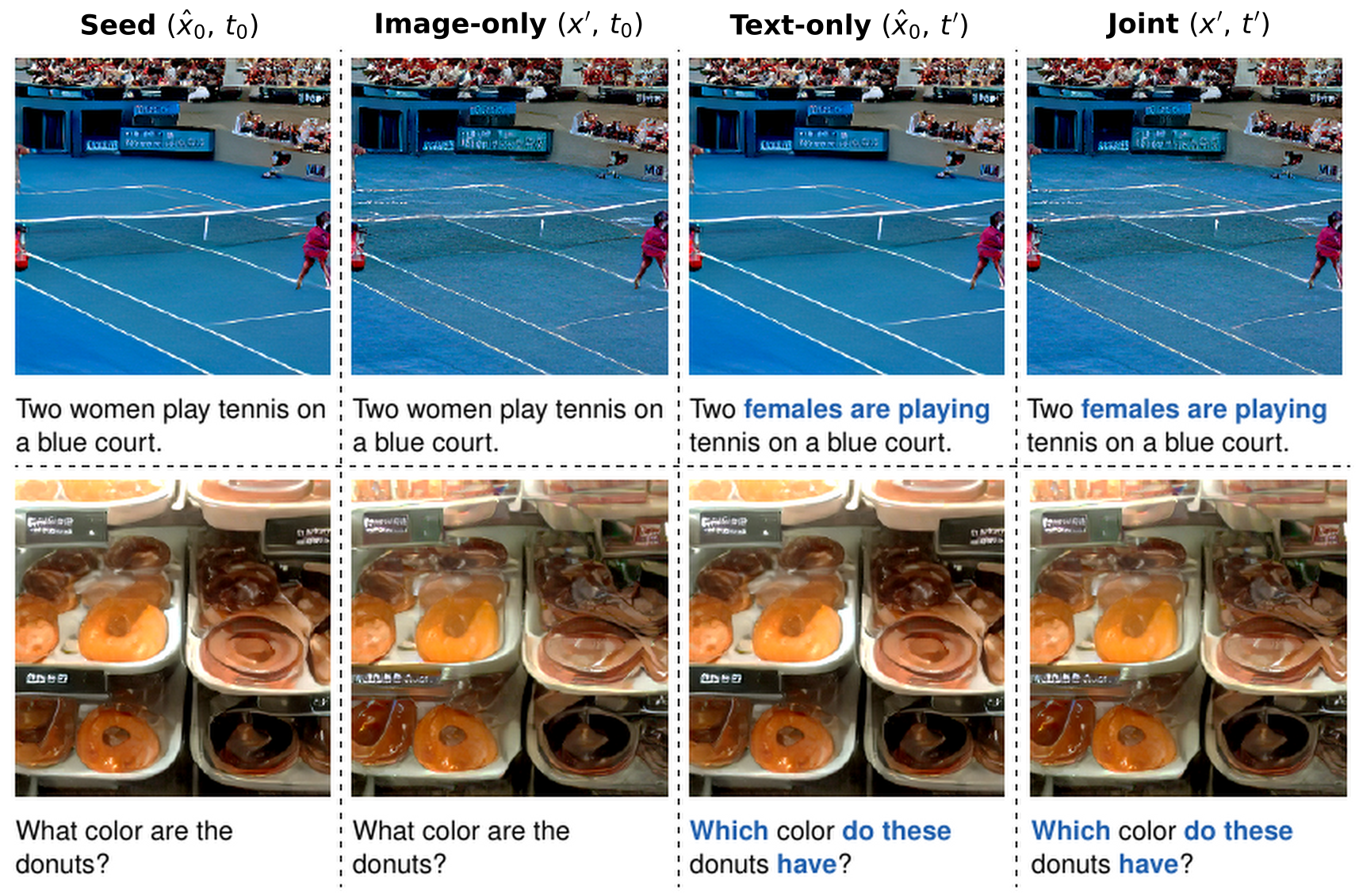}
\caption{RQ3: two attributed joint failures, each with the reconstructed seed $(\hat x_0,t_0)$, the image-only input, the text-only input, and the joint input. Changed words are in bold. Model outputs from left to right: match, match, match, no match (ITM); brown, brown, brown, yellow (VQA).}
\label{fig:case}
\end{figure*}

Figure~\ref{fig:case} shows two attributed joint failures with their four cells. In the ITM case, the image variant changes the shade and texture of the court, and the text variant rewrites ``Two women play tennis on a blue court'' as ``Two females are playing tennis on a blue court''. The model predicts a match for the seed and both single-modality inputs and no match for the joint input, although the caption still describes the image. In the VQA case, the image variant shifts the shading of the donuts, and the text variant rewrites ``What color are the donuts?'' as ``Which color do these donuts have?''. The model answers ``brown'' for the seed and both single-modality inputs and ``yellow'' for the joint input. Single-modality testing passes both cases.

\begin{table*}[t]
\centering
\footnotesize
\caption{RQ3: types of attributed joint failures in a stratified random sample of 180, 60 per task.}
\label{tab:taxonomy}
\renewcommand{\arraystretch}{0.82}
\setlength{\tabcolsep}{3pt}
\begin{tabular*}{\textwidth}{@{\extracolsep{\fill}}lrrrr}
\toprule
\textbf{Failure type} & \textbf{ITM} & \textbf{VR} & \textbf{VQA} & \textbf{Total} \\
\midrule
Attribute or entity binding & 24 & 9 & 19 & 52 \\
Spatial or relational composition & 17 & 15 & 14 & 46 \\
Quantifier, count, or scope & 4 & 26 & 4 & 34 \\
Lexical grounding or co-reference & 11 & 6 & 12 & 29 \\
Fine-grained category or part & 4 & 4 & 11 & 19 \\
\midrule
Total & 60 & 60 & 60 & 180 \\
\bottomrule
\end{tabular*}
\end{table*}

\subsection{RQ4: Sensitivity and Generalization}
\label{sec:rq4}

\subsubsection{Design Choices}
We vary one design choice of \tool\ (SF) at a time on BLIP (ITM), BEiT-3 (VR), and PaliGemma (VQA), with the same 100 seeds, budget, and other settings: (a) the relation filters, with no filter, each filter alone, or all filters; (b) the pool size $n\in\{250,500,750\}$, which sets the single-modality calls to $2n$ while the joint calls stay at $B$; (c) the rank range $R\in\{2{,}047,4{,}095,8{,}191\}$; (d) the number of edited indices $\kappa\in\{5,10,15,20,30\}$; and (e) the text pipeline with rewrite, expand, or prune removed. These are 16 settings and 4,800 runs.

Table~\ref{tab:ablation} reports averages over the three pairs; $\cfk$ is computed over the single-modality calls of each setting, and the default appears once in every group. The default reaches a confirmation rate of 81.6\%, 55.1 $\cfk$, and 145.9 $\vjfk$. Without filters, \tool\ (SF) finds 25\% fewer attributed joint failures (108.7) and its confirmation rate drops to 71.8\%. Among the single filters, $F_{\mathrm{pair}}$ achieves the highest $\vjfk$ (131.4), followed by $F_{\mathrm{txt}}$ (123.5) and $F_{\mathrm{img}}$ (120.2). However, its overall candidate confirmation rate (71.9\%) remains close to that of the unfiltered setting (71.8\%). These metrics capture different aspects of testing effectiveness: the confirmation rate measures the proportion of candidates confirmed as failures across input types, whereas $\vjfk$ measures attributed joint failures per 1,000 joint calls. A pool of 250 variants finds 9\% fewer than the default of 500, and a pool of 750 about as many (147.0). Narrowing the rank range reduces $\vjfk$ to 136.7 at $R=4{,}095$ and 121.4 at $R=2{,}047$, while the confirmation rate remains between 80.8\% and 81.3\%. $\vjfk$ peaks at $\kappa=15$ and $20$ (145.9 and 146.6) and drops at $\kappa=30$ (140.7), where $\cfk$ also falls to 51.8. Removing any text stage lowers the confirmation rate to 73.9\% to 79.6\% and $\vjfk$ by 9\% to 13\%; removing the prune stage lowers $\cfk$ the most, to 47.5.

\begin{table*}[t]
\centering
\footnotesize
\caption{RQ4: design choices of \tool\ (SF), averaged over BLIP (ITM), BEiT-3 (VR), and PaliGemma (VQA).}
\label{tab:ablation}
\renewcommand{\arraystretch}{0.82}
\setlength{\tabcolsep}{4pt}
\begin{tabular*}{\textwidth}{@{\extracolsep{\fill}}llrrr}
\toprule
\textbf{Design choice} & \textbf{Setting} &
\makecell{\textbf{Conf.}\\\textbf{rate (\%)}} & $\cfk$ & $\vjfk$ \\
\midrule
\multirow[t]{5}{*}{(a) relation filters}
 & none                       & 71.8 & 46.2 & 108.7 \\
 & $F_{\mathrm{txt}}$ only    & 77.6 & 51.3 & 123.5 \\
 & $F_{\mathrm{img}}$ only    & 76.4 & 50.6 & 120.2 \\
 & $F_{\mathrm{pair}}$ only   & 71.9 & 52.8 & 131.4 \\
 & all (default)              & \textbf{81.6} & \textbf{55.1} & \textbf{145.9} \\
\midrule
\multirow[t]{3}{*}{(b) pool size}
 & $n=250$                    & 80.7 & 51.8 & 132.6 \\
 & $n=500$ (default)          & \textbf{81.6} & \textbf{55.1} & 145.9 \\
 & $n=750$                    & 81.3 & 54.9 & \textbf{147.0} \\
\midrule
\multirow[t]{3}{*}{(c) rank range ($\kappa=15$)}
 & $R=2{,}047$                & {81.3} & 48.6 & 121.4 \\
 & $R=4{,}095$                & 80.8 & 52.9 & 136.7 \\
 & $R=8{,}191$ (default)      & \textbf{81.6} & \textbf{55.1} & \textbf{145.9} \\
\midrule
\multirow[t]{5}{*}{(d) edited indices ($R=8{,}191$)}
 & $\kappa=5$                 & {80.1} & 47.3 & 118.8 \\
 & $\kappa=10$                & 80.8 & 52.6 & 135.2 \\
 & $\kappa=15$ (default)      & 81.6 & {\textbf{55.1}} & {145.9} \\
 & $\kappa=20$                & 81.2 & 55.0 & \textbf{146.6} \\
 & $\kappa=30$                & \textbf{81.9} & 51.8 & 140.7 \\
\midrule
\multirow[t]{4}{*}{(e) text stages}
 & $-$rewrite                 & 73.9 & 51.1 & 126.5 \\
 & $-$expand                  & {79.6} & 52.7 & 132.9 \\
 & $-$prune                   & 75.8 & 47.5 & 128.4 \\
 & full (default)             & \textbf{81.6} & \textbf{55.1} & \textbf{145.9} \\
\bottomrule
\end{tabular*}
\end{table*}

\subsubsection{General-Purpose Models}
We also test four general-purpose VLMs on all three tasks: Qwen3-VL-8B-Instruct and Qwen3-VL-32B-Instruct~\cite{qwen2025qwen3vl}, one open-weight family at two scales, and two commercial models, GPT-4o mini~\cite{openai2024gpt4omini} in its snapshot of 18 July 2024 and Gemini 2.5 Flash~\cite{comanici2025gemini}. On each model, we run \tool\ (SF), SR, and CRND. The Qwen3-VL models use the settings of Section~\ref{sec:budget} with 100 seeds per task. The API models use 50 sampled seeds per task, $B=50$, and $n=25$; Table~\ref{tab:modern} lists the admitted seeds $|S|$.
For ITM and VR, the prompts and the rules for extracting a label are fixed in advance, and an output that maps to no label counts as an abstention. The API models are called at temperature~0 where the API allows it, and every call is logged with its date. The API experiments use 37,650 single-modality calls and 37,650 joint calls for failure discovery.

Table~\ref{tab:modern} shows the results. \tool\ (SF) finds 111.8 to 249.8 attributed joint failures per 1,000 joint calls, and its rate exceeds that of SR (by 5.6 to 24.1) and CRND (by 21.6 to 41.3) on all twelve task--model pairs. On the API models, the differences rest on few failures: the smallest, 5.6 for GPT-4o mini on ITM, corresponds to about 12 attributed joint failures. As on the task-specific models, VQA yields the most attributed joint failures and VR the fewest. Qwen3-VL-32B yields 19.8\% to 29.4\% fewer attributed joint failures than Qwen3-VL-8B on every task, each on its own admitted seeds. On the API models, 2.3\% to 4.7\% of the admitted seeds show output variation.

\begin{table*}[t]
\centering
\footnotesize
\caption{RQ4: \tool\ (SF) on the general-purpose models. $|S|$: admitted seeds. $\Delta$ vs.\ SR and $\Delta$ vs.\ CRND: differences in $\vjfk$. Non-det.\ floor: share of admitted seeds with observed output variation. }
\label{tab:modern}
\renewcommand{\arraystretch}{0.82}
\setlength{\tabcolsep}{3pt}
\begin{tabular*}{\textwidth}{@{\extracolsep{\fill}}llrcccc}
\toprule
\textbf{Task} &
\textbf{Model} &
$|S|$ &
\makecell{\textbf{Non-det.}\\\textbf{floor}} &
$\vjfk$ &
$\Delta$ vs.\ SR &
$\Delta$ vs.\ CRND \\
\midrule
ITM & Qwen3-VL-8B & 100 & -- & 169.0 & 18.2 & 33.7 \\
ITM & Qwen3-VL-32B & 100 & -- & 135.6 & 14.4 & 29.8 \\
ITM & GPT-4o mini & 43 & 4.7\% & 139.5 & 5.6 & 33.0 \\
ITM & Gemini 2.5 Flash & 41 & 2.4\% & 127.3 & 20.0 & 27.8 \\
\midrule
VR & Qwen3-VL-8B & 100 & -- & 148.2 & 21.0 & 28.2 \\
VR & Qwen3-VL-32B & 100 & -- & 111.8 & 13.2 & 25.0 \\
VR & GPT-4o mini & 38 & 2.6\% & 121.6 & 7.4 & 21.6 \\
VR & Gemini 2.5 Flash & 40 & 2.5\% & 125.0 & 17.5 & 24.5 \\
\midrule
VQA & Qwen3-VL-8B & 100 & -- & 249.8 & 19.2 & 37.4 \\
VQA & Qwen3-VL-32B & 100 & -- & 176.4 & 14.8 & 32.2 \\
VQA & GPT-4o mini & 45 & 4.4\% & 172.9 & 12.9 & 41.3 \\
VQA & Gemini 2.5 Flash & 44 & 2.3\% & 193.6 & 24.1 & 32.7 \\
\bottomrule
\end{tabular*}
\end{table*}

\section{Discussion}
\label{sec:discussion}

\subsection{Output Changes Overstate Failures}
Table~\ref{tab:rq1} gives the share of candidates that are confirmed failures, by input type: 86.3\% to 91.6\% for image-only inputs, 77.0\% to 86.8\% for text-only inputs, and 75.0\% to 83.0\% for joint inputs. A count of output changes therefore overstates the failures of \tool\ (SF) by 9\% to 16\% (relative) on image-only inputs, 15\% to 30\% on text-only inputs, and 20\% to 33\% on joint inputs; on VQA it also counts the 6.4\% of output changes that stay within the reference answers. The overstatement depends on the generator: without the relation filters, the confirmation rate falls from 81.6\% to 71.8\% (Table~\ref{tab:ablation}), so a count of output changes would credit the unfiltered generators with 39\% more failures than they find, against 23\% for the filtered ones. Counting output changes is still useful for screening, but the effectiveness of a generator should be measured on confirmed failures.

\subsection{What Four-Cell Evidence Adds}
A count of joint failures without single-modality evidence mixes failures caused by the image mutation, failures caused by the text mutation, and failures caused by their combination. Four-cell confirmation rejects 22.5\% of the joint candidates of \tool\ (SF) (Table~\ref{tab:fourcell}), but most of these rejections, 19.7\% of the joint candidates, come from checking the joint input itself; the single-modality controls reject a further 2.8\%. The main contribution of the controls is therefore attribution: for every reported joint failure, the four cells provide evidence that the model handles each mutation alone but fails on their combination. The interaction contrast then separates the 16.5\% of score-classifiable attributed joint failures that are interactions between the modalities from the rest, where two single-modality effects cross the decision boundary only together (Table~\ref{tab:interaction}). These additive failures are still failures on valid inputs that single-modality testing passes, and the positive median $\Delta_{\mathrm{int}}$ shows that, within these failures, the two effects usually cancel in part rather than reinforce each other.

\subsection{Implications}
For VLM developers, more than half of the sampled attributed joint failures concern attribute binding or spatial composition, and quantifier and count errors cluster on VR (Table~\ref{tab:taxonomy}). Each attributed joint failure comes with its four cells and can serve as a regression test that checks the two mutations alone and together. The rates of attributed joint failures should not be read as a ranking of tasks or models: they depend on how often the oracle rejects candidates, which differs between tasks (Table~\ref{tab:fourcell}), and on the admitted seeds, which differ between models. For testing researchers, checking the expected answer removes a fifth of the counted failures, and matched single-modality controls are needed to attribute a joint failure to the combination; both belong in the evaluation of multimodal testing studies.

\section{Threats to Validity}
\label{sec:threats}

\noindent\textbf{Internal Validity.}
The oracle relies on a VLM judge, which can make mistakes. RQ1 measures the precision of the failures it reports for every method and task and checks a sample of the joint candidates of \tool\ (SF) that it rejects; we do not measure how many failures it misses for the other methods. The judge receives the reference answers as a reference, which may bias its answer check toward preserved; missed answer changes are among the oracle errors that the annotators find (Section~\ref{sec:rq1}). Under the joint-input-first attribution of rejection reasons, the single-modality controls reject a further 2.8\% of joint candidates; these rejections include cases with conflicting judge verdicts and are not evidence that the model fails on the single-modality input. All cells are built on the reconstructed seed, and seeds whose reconstruction changes the expected answer are not admitted. Admitting seeds per model favors inputs that the model handles correctly, and we report the admission rates. The filters and the judge share no model family or vision encoder with the models under test.
All counts rest on the oracle verdicts, whose precision RQ1 measures. $\cfk$ and $\vjfk$ count failing inputs, not distinct faults; VarCov shows how many distinct variants the attributed joint failures involve.

\noindent\textbf{External Validity.}
Our study covers three tasks, eight models, and one tokenizer family. All cells are $256\times256$ VQGAN reconstructions, which carry reconstruction artifacts; the oracle checks that each reported input is a valid input for the task, and the annotators check this on a sample. Results may differ for other tokenizers and generators, for high-resolution or text-heavy images, and for multi-turn systems. The seeds come from public validation splits that the models under test and the judge may have seen during training; the generated variants are new inputs, but they stay close to these seeds. The results for the API models are tied to the model versions available during the study and to a reduced budget.

\section{Conclusion}
\label{sec:conclusion}

This paper presented \tool, a black-box approach for testing VLMs that finds failures whose expected answers are confirmed by an oracle and attributes joint failures to the combination of image and text mutations with four-cell evidence. \tool\ combines controlled visual and textual generators, relation filters, and two-phase testing with feedback-directed pairing. An automated oracle confirms every reported failure, and human labels on a stratified sample measure its precision. The oracle confirms 81.7\% of the candidates of \tool\ (SF) and agrees with human labels on 80.0\% to 94.6\% of the failures it reports across the compared methods in the primary experiment, and on 88.8\% to 94.6\% of those reported by the default configuration of \tool. Under the same budget, the generators of \tool\ find 1.8 to 2.5 times as many confirmed failures as SINVAD and TextFooler, and \tool\ finds 91 to 197 attributed joint failures per 1,000 joint calls, 1.3 times as many as the same pairing applied to the variants of these two baselines. Four-cell confirmation rejects 22.5\% of the joint candidates, 19.7\% through the joint input alone and 2.8\% through the single-modality controls. Under an additive margin model, 16.5\% of the score-classifiable attributed joint failures are interactions between the modalities; in the rest, the additive prediction alone already fails, although each single-modality input passes. The relation filters account for most of the difference between configurations, and on four general-purpose models, including two commercial API models, \tool\ (SF) finds more attributed joint failures than CRND and SR on every task.

\bibliographystyle{ACM-Reference-Format}
\bibliography{acm}

\end{document}